\documentclass[aps,prb,twocolumn,superscriptaddress,showpacs]{revtex4}
\usepackage{graphicx}
\usepackage{latexsym}
\usepackage{amssymb}
\usepackage{amsmath}
\usepackage{amsfonts}
\usepackage{bm}
\usepackage{multirow}
\usepackage{color}
\usepackage{mathbbol}
\usepackage{bm}
\newcommand{\ii}{\mathrm{i}}
\newcommand{\dd}{\mathrm{d}}

\newcommand{\SU}{\mathrm{SU}}
\newcommand{\U}{\mathrm{U}}

\newcommand{\beq}{\begin{equation}}
\newcommand{\eeq}{\end{equation}}
\newcommand{\beqn}{\begin{eqnarray}}
\newcommand{\eeqn}{\end{eqnarray}}
\newcommand{\nn}{\nonumber}

\DeclareMathAlphabet{\mathbbold}{U}{bbold}{m}{n}

\def\sgn{{\rm sgn}}

\def\SU{{\rm SU}}

\def\U{{\rm U}}

\def\mr{\mathrm{m}}

\usepackage{ulem}
\newcommand{\dg}{\dagger}

\newcommand{\bs}{\boldsymbol}

\begin{document}

\title{A construction of exotic metallic states}

\author{Xiao-Chuan Wu}
\affiliation{Department of Physics, University of California,
Santa Barbara, CA 93106, USA}

\author{Yichen Xu}
\affiliation{Department of Physics, University of California,
Santa Barbara, CA 93106, USA}

\author{Mengxing Ye}
\affiliation{Kavli Institute for Theoretical Physics, University
of California, Santa Barbara, CA 93106, USA}

\author{Zhu-Xi Luo}
\affiliation{Kavli Institute for Theoretical Physics, University
of California, Santa Barbara, CA 93106, USA}

\author{Cenke Xu}
\affiliation{Department of Physics, University of California,
Santa Barbara, CA 93106, USA}

\begin{abstract}

We discuss examples of two dimensional metallic states with charge
fractionalization, and we will demonstrate that the mechanism of
charge fractionalization leads to exotic metallic behaviors at low
and intermediate temperature. The simplest example of such state
is constructed by fermionic partons at finite density coupled to a
$Z_N$ gauge field, whose properties can be studied through
rudimentary methods. This simple state has the following exotic
features: (1) at low temperature this state is a ``bad metal"
whose resistivity can be much larger than the Mott-Ioffe-Regel
limit; (2) while increasing temperature $T$ the resistivity
$\rho(T)$ is a nonmonotonic function, and it crosses over from a
bad metal at low $T$ to a good metal at relatively high $T$; (3)
the optical conductivity $\sigma(\omega)$ has a small Drude weight
at low $T$, and a larger Drude weight at intermediate $T$; (4) at
low temperature the metallic state has a large Lorenz number,
which strongly violates the Wiedemann-Franz law. A more complex
example with fermionic partons at finite density coupled to a
$\SU(N)$ gauge field will also be constructed.

\end{abstract}

\maketitle


\section{Introduction}

Dimensionless quantities in nature can be universal, meaning they
are insensitive to the microscopic details of the system.
Dimensionless universal quantities can arise from two different
mechanisms: either criticality, or topology. At a critical point
(either classical or quantum critical point), the diverging
correlation length renders most of the the microscopic details
irrelevant to infrared physics, hence each universality class is
characterized by a series of numbers referred to as critical
exponents. Examples of these critical points include various two
dimensional statistical mechanics models such as the Ising
model~\cite{yangising}, the ``Yang-Lee
singularity"~\cite{yanglee}, and the Wilson-Fisher fixed points of
three dimensional systems~\cite{wilsonfisher}. Topology can lead
to universal quantities due to topological quantization. The
simplest example of such is the magnetic flux quantization in
Dirac monopole~\cite{dirac}, and in
superconductor~\cite{fluxquantum1,fluxquantum2}. The Hall
conductivity of quantum Hall systems (either integer or
fractional) is a discrete universal number, it is related to the
level of the Chern-Simons topological field
theory~\cite{zhangcs,wencs,zhangcs2,wencs2}, which has to be
quantized due to mathematical consistency.

Electrical resistivity/conductivity is a dimensionless quantity in
two spatial dimensions, hence it can in principle take universal
values that are independent of the microscopic details of the
system. A universal resistivity can arise with various mechanisms.
Besides the Hall resistivity of the quantum Hall states mentioned
above, the resistivity of $(2+1)d$ quantum critical points with
gapless charge degree of freedom~\cite{UC1,UCt1}, the resistivity
jump at $(2+1)d$ metal-insulator transition driven by
interaction~\cite{senthilmit1,senthilmit2,resistivity2}, and the
criterion of the so-called ``bad metal" in two
dimensions~\cite{limit1,limit2,limit3} are all ``universal". In
all these examples, the resistivity (or the bound of resistivity)
is always an order-unity dimensionless number times $h/e^2$.

This work concerns the metallic states with finite charge density
and finite charge compressibility. The usual theory that describes
the transport of a metal
is the Boltzmann equation. 
The Boltzmann equation most conveniently applies when the concept
of quasiparticles remains valid in the system~\footnote{A
generalized quantum Boltzmann equation can be developed when
well-defined quasiparticles are lost due to interaction with
bosonic modes~\cite{kim}.}, which usually requires that
$l_\mathrm{m} k_F \gg 1$, where $l_\mr$ is the mean free path, and
$k_F$ is the Fermi wave vector. When $l_\mathrm{m} k_F$ becomes
order 1, the resistivity saturates the Mott-Ioffe-Regel (MIR)
limit of a metal, and the system becomes a ``bad
metal"~\cite{limit1,limit2,limit3}, where descriptions based on
quasiparticles break down. For a purely two dimensional system,
the condition of $l_\mathrm{m} k_F \sim l_{\mathrm{m}} n^{1/2}
\sim 1$ implies that the resistivity $\rho$ should be at the order
of $h/e^2$. When the measured resistivity $\rho$ of a purely two
dimensional metal is significantly larger than $h/e^2$, or in
other words the estimated value of $l_\mathrm{m} k_F$ exceeds
order unity for a $2d$ metal, one has to abandon the conventional
description based on quasiparticles, and resort to other
theoretical tools.

In real systems metallic states without quasiparticles usually
arise from coupling electrons to bosonic gapless quantum critical
modes. The theoretical formalism for these states usually start
with a decoupled system with noninteracting electrons, and analyze
how the fermion-boson coupling modifies the
system~\cite{polchinskinfl,nayaknfl1,nayaknfl2,nfl2,Mross2010,nfl4,nfl5}.
Through various perturbative renormalization group methods, one
can show that the coupling between the Fermi surface and the
gapless bosonic modes is relevant, and potentially drive the
system into a non-Fermi liquid fixed point without quasiparticles.
In recent years, a new route of constructing non-Fermi liquid has
been explored, which was based on models that are soluble in
certain limit (such as the Sachdev-Ye-Kitaev model and other
related
models)~\cite{SachdevYe1993,Kitaev2015,MaldacenaStanford2016,Witten2016,Klebanov2016}.
These models have no notion of spatial dimensions, but solution of
these models already have no quasiparticles. Lattice models built
upon these soluble models quite naturally lead to non-Fermi
liquids in various spatial
dimensions~\cite{Song2017,phonon1,phonon2,patel2017,patel2018,berg2018,xu2018,wu2019}.

In this work we explore an alternative construction of exotic
metallic states. The constructions used in this work are not based
on soluble lattice models of interacting electrons, but there are
sufficient theoretical arguments to show that these are indeed
stable states. Though these examples are far from weakly
interacting electrons with quasiparticles, the design of these
states allows them to be studied through rudimentary theoretical
tools.

\section{$Z_N$ fractionalized metal}

The central idea of our construction is ``charge
fractionalization". Fractionalization of quantum numbers is most
well-known and well established in particle
physics~\cite{gellmann}, but it is also predicted and observed in
condensed matter systems such as fractional quantum Hall
states~\cite{laughlin,fraccharge1,fraccharge2}. Quantum number
fractionalization is also one of the signatory phenomena in
quantum spin
liquids~\cite{anderson,laughlinspinliquid,wenspinliquid1,subirspinliquid,wenspinliquid2}.
Electric charge fractionalization was discussed in the context of
Mott transition in systems with partially filled $3d$ pyrochlore
lattice~\cite{gang2014}. Recently, motivated by experiments on
transition metal dichalcogenide (TMD) moir\'{e}
heterostructures~\cite{tmdmit1}, effects of charge
fractionalization at the metal-insulator transition in pure $2d$
systems have been discussed in
Ref.~\onlinecite{xumit,debanjan2021}. In this work we will explore
the consequences of charge fractionalization in a metallic state.

The first example we consider is a $Z_N$ topological order
enriched with a global $\U(1)$ symmetry, which corresponds to the
ordinary electric charge conservation. The elementary anyon
$f_\alpha$ of the $Z_N$ topological order carries a $Z_N$ gauge
charge, and it is also a spin-1/2 fermion with fractional electric
charge $e_* = e/N$. When $N$ is an odd integer, the gauge
invariant states of the system include fermions that carry odd
integer electric charges and half-integer spins; as well as bosons
with even integer charges and integer spins. This is the same
Hilbert space as a many-body electron system.

It is known that the discrete gauge field at two spatial
dimensions has a stable deconfined phase at zero temperature, in
which the anyon $f_\alpha$ can be separated infinitely far from
each other, hence the anyon $f_\alpha$ plays the role as the
charge carriers in the system at least at zero temperature.
Although we do not pursue an exactly soluble model based on
electrons in this work, the state discussed here should be a
stable state of electrons, given that a discrete gauge theory is
free from confinement at zero temperature in $(2+1)d$. At finite
temperature the thermal equilibrium state of the system is in a
confined phase, but at low temperature the observable physics
should still crossover to the deconfined phase at zero
temperature. In fact, the finite temperature confinement of the
$Z_N$ gauge field is caused by thermally activated population of
gauge fluxes with nontrivial mutual statistics with $f_\alpha$.
The confinement length $\xi(T)$, i.e. the distance that a single
$f_\alpha$ can be separated from the ``crowd", takes the form of
$\xi(T) \sim \exp(c \Delta/T)$, where $\Delta$ is the gap of $Z_N$
gauge fluxes, and $c$ is a constant. We argue that when $\xi(T)$
simultaneously satisfy two criteria, namely ({\it i.}) $\xi(T)$ is
large compared with the distance between $f_\alpha$ anyons
(assuming a sufficiently large charge density), and ({\it ii.})
$\xi(T)$ is large compared with the mean free path $l_\mathrm{m}$,
the anyon $f_\alpha$ still plays the role of charge carrier in the
nonequilibrium process of charge transport, as $f_\alpha$ does not
travel long enough between two consecutive scatterings to ``feel"
the confinement.

$ $

{\it --- Charge Transport}

In the follows we will discuss various properties of the state
described above. We first consider electrical resistivity at zero
temperature. The key advantage of this construction is that, at
zero temperature, the $Z_N$ gauge field dynamics is gapped, and
does not lead to any scattering to the gapless charged partons
below the gap of the $Z_N$ gauge fields. The main source of
relaxation of electric current at zero temperature still comes
from conventional mechanisms, such as impurities, which give the
partons a mean free length $l_\mathrm{m}$. If we assume the
electric charge density is $e n_e$, in an ordinary system without
fractionalization, the rudimentary semiclassical theory of
transport breaks down when $l_\mathrm{m} n_e^{1/2} \sim 1$, i.e.
$l_\mathrm{m} \sim 1/n_e^{1/2}$. Since the parton carries charge
$e/N$, the density of the parton is $n_\ast = N n_e $, hence the
usual transport theory can be applicable for even smaller
$l_\mathrm{m}$, i.e. $l_\mathrm{m} \sim 1/(n_\ast)^{1/2} \sim 1/(N
n_e)^{1/2}$. When $l_\mathrm{m}$ saturates this limit, the system
should be a ``bad metal of partons", with an ``upper bound" of
resistivity \beqn \rho_{max} \sim \frac{h}{e_*^2} \sim N^2
\frac{h}{e^2}. \eeqn When $l_\mathrm{m} \sim 1/(n_\ast)^{1/2}
> 1$, rudimentary formalism of describing metallic states should
still apply; the only difference is that now the charge carrier
$f_\alpha$ carries charge $e_\ast = e/N$, and the density of
$f_\alpha$ is higher than electric charge density.

At low temperature, a $2d$ $Z_N$ gauge field will cause
confinement in equilibrium. As we mentioned above, the confinement
of a $2d$ $Z_N$ gauge theory is caused by the thermally activated
gauge fluxes, and the confinement length $\xi(T)$ is roughly the
distance between two thermally activated gauge fluxes, hence
$\xi(T) \sim \exp(c\Delta/T)$, where $\Delta$ is the energy gap
for the $Z_N$ gauge flux. We need to compare $\xi(T)$ with other
length scales of the system: the distance between anyons
$f_\alpha$, which is given by $1/n_\ast^{1/2}$; the lattice
constant $a$; and the mean free path $l_\mathrm{m}$. To ensure
that the transport of the state can be studied with controlled
methods, we assume that the mean free length $l_\mathrm{m}$ is at
the order of, or larger than $N^{1/2}/ n_e^{1/2}$; or equivalently
$l_\mr n_\ast^{1/2}$ is at the order of, or greater than $N$. In
this limit, at least at low temperature, the following hierarchy
of length scales holds: $\xi(T) > l_\mathrm{m} > 1/n_\ast^{1/2}$.
In this limit the simple theory of metal, such as the Drude
formula still applies. The conductivity at zero and low
temperature would be \beqn \sigma_0 = \frac{n_\ast e_\ast^2
l_\mr}{m_\ast v^\ast_F} \sim \frac{e_\ast^2}{h} (l_\mr
n_\ast^{1/2}) \sim \frac{1}{N} \frac{e^2}{h}, \eeqn which can
still be a bad metal, even with the choice of relatively long mean
free length.

We assume that $l_\mathrm{m}$ mostly arises from scattering with
impurities with a hard-sphere like potential, and hence is
insensitive to temperature. With rising temperature, the
resistivity first increases with conventional mechanism, such as
scattering with phonon, or interaction between the partons. These
scattering are still suppressed due to the small electric charge
carried by the partons. For example, the parton-phonon interaction
is down by a factor of $1/N$ compared with the electron-phonon
interaction, and the resistivity due to parton-phonon interaction
is down by a factor of $1/N^2$. For short-range parton-parton
interactions which presumably leads to Fermi liquid like scaling
of resistivity (i.e. $\rho \sim \rho_0 + A T^2$), if the
short-range interaction arises from screened Coulomb interaction,
the interaction is suppressed by a factor of $1/N^2$~\footnote{The
screening of the Coulomb interaction is affected by the charge of
the parton as well, which will complicate the estimate of the
screened-short range interaction.}.

\begin{figure}
\includegraphics[width=0.4\textwidth]{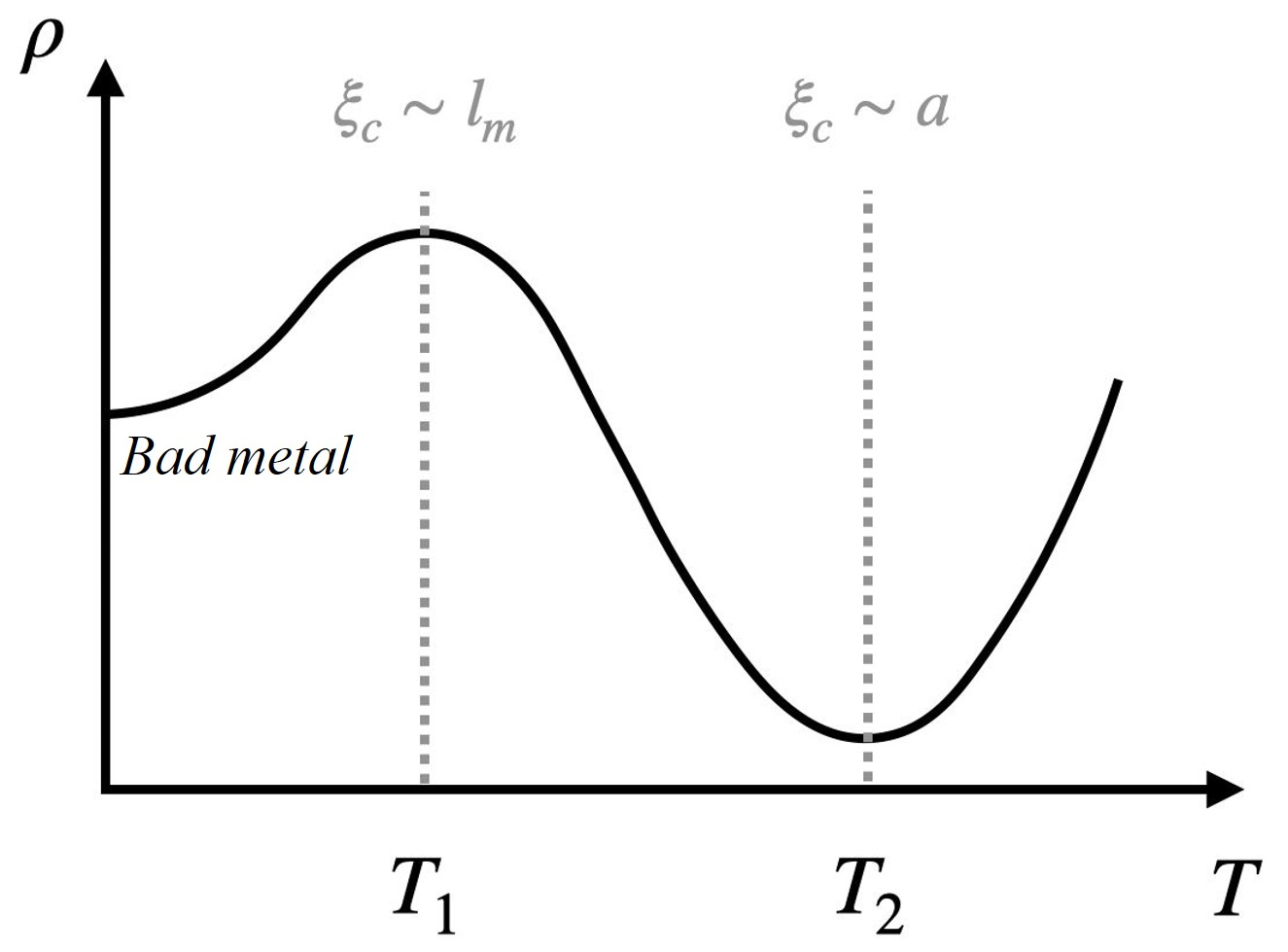}
\caption{The schematic behavior of resistivity $\rho(T)$
constructed with fermionic partons carrying fractional charges
coupled with a $Z_N$ gauge field. } \label{resistivity}
\end{figure}

When temperature rises further, the confinement length $\xi(T)$
becomes shorter, and eventually at temperature scale $T_1$ where
$\xi(T_1)\sim l_\mathrm{m}$, the semiclassical picture of
$f_\alpha$ breaks down. At even higher temperature scale where the
confinement length $\xi(T)$ is comparable with the lattice
constant $a$, i.e. $T > T_2$ with $\xi(T_2) \sim a$, the partons
are fully confined, and the charge carriers should still be viewed
as electrons. The electron density is $n_e$, and since we assumed
a hard-sphere like potential of the impurities, $l_\mr$ from
impurities remains approximately unchanged from before. The
conductivity at temperature $T_2$ should be \beqn \sigma(T_2) \sim
\frac{n_e e^2 l_\mr}{m v_F} \sim N^{1/2} \frac{e^2}{h}, \eeqn
which can be a good metal. Hence with rising temperature, the
resistivity evolves in a nonmonotonic way; it will crossover from
a bad metal with $T < T_1$ to a good metal at $T \sim T_2$. The
schematic behavior of $\rho(T)$ is sketched in
Fig.~\ref{resistivity}.

We have chosen $l_\mr$ so that the simple pictures of metal such
as the Drude theory applies for both temperature ranges $T < T_1$
and $T > T_2$. In the low temperature range the semiclassical
theory of metal with fractional charge carrier $f_\alpha$ becomes
applicable; while with $T > T_2$ the system becomes a conventional
metal with electrons. We lack the reliable theoretical tools to
describe the intermediate temperature range $T_1 < T < T_2$, but
sufficient argument can lead to the conclusion that the system
crossover from a bad metal phase at low temperature range, to a
good metal phase in the higher temperature range. Also, if the
optical conductivity is measured, our construction implies that
the Drude weight of the optical conductivity is small at $T <
T_1$, but the Drude weight will crossover to a larger value
proportional to $\sigma(T_2)$ at $T \sim T_2$.

{\it --- Hall Effect}

For both temperature ranges $T < T_1$ and $T > T_2$, the transport
coefficients can be derived with the rudimentary semiclassical
theory of metal. We take the semiclassical Boltzmann transport
equation under the relaxation-time approximation \beqn
\frac{\partial g}{\partial
t}+\dot{\boldsymbol{x}}\cdot\frac{\partial
g}{\partial\boldsymbol{x}}+\dot{\boldsymbol{k}}\cdot\frac{\partial
g}{\partial\boldsymbol{k}}=\left(\frac{\partial g}{\partial
t}\right)_{\textrm{coll}}\approx-\frac{\delta g}{\tau}, \label{eq:
Boltzmann eq} \eeqn where $g(t,\boldsymbol{x},\boldsymbol{k})$
denotes the non-equilibrium distribution function, and $\delta g$
is its deviation from the equilibrium distribution $f(\epsilon)$.
This Boltzman equation can be applied to the parton $f_\alpha$ at
temperature $T < T_1$, and to electrons at temperature $T > T_2$.

At low temperature, according to the Ong's formula~\cite{ong}
based on the Jones-Zener solution to Eq.~\ref{eq: Boltzmann eq},
the weak-field Hall conductivity in $2d$ metals has a geometric
interpretation \beqn
\sigma_{xy}=\frac{e_{*}^{2}}{h}\frac{\mathcal{A}_{l}}{\pi(l_{B}^{*})^{2}}
\eeqn where $l_{B}^{*}=\sqrt{\hbar/(e_{*}B)}$ is the magnetic
length for partons, and $\mathcal{A}_{l}$ is the area swept out by
the vector
$\boldsymbol{l}(\boldsymbol{k})=\tau(\epsilon(\boldsymbol{k}))\boldsymbol{v}(\boldsymbol{k})$
when $\boldsymbol{k}$ moves around the FS, i.e., \beqn
\mathcal{A}_{l}=\frac{\boldsymbol{B}}{B}\cdot\int_{\textrm{FS}}d\boldsymbol{l}(\boldsymbol{k})\times\boldsymbol{l}(\boldsymbol{k})\sim
l_{\textrm{m}}^{2}. \eeqn
If we assume electrons and partons share the same isotropic
$l_{\textrm{m}}$ and therefore the same area $\mathcal{A}_{l}$,
there is a large ratio between $\sigma_{xy}$ at low temperature
and the second crossover temperature $T_2$: \beqn
\frac{\sigma_{xy}(T_{2})}{\sigma_{xy}(T < T_1)} \sim N^{3}. \eeqn

The situation is different in the strong field limit. In this
limit the collision integral in Eq.~\ref{eq: Boltzmann eq} can be
neglected. Taking the directions $\boldsymbol{B}=B\hat{z}$ and
$\boldsymbol{E}=E\hat{y}$, one obtains the solution $\delta
g=(\hbar k_{x}E/B)(\partial f/\partial\epsilon)$ which has no
explicit dependence on $N$. In this case, the integral of
$v_{x}(\boldsymbol{k})\delta g(\boldsymbol{k})$ over the Brillouin
zone gives $n_{*}E/B$, which leads to the high-field Hall
conductivity \beqn \sigma_{xy}=\frac{n_\ast
e_\ast}{B}=\frac{n_{e}e}{B}. \eeqn The answer only depends on the
total charge density.

$ $

{\it --- Thermoelectric Properties}

In the presence of nonzero electric field $\boldsymbol{E}$ and
temperature gradient $-\nabla T$, the linear response of electric
current $\boldsymbol{J}^{e}$ and heat current $\boldsymbol{J}^{h}$
are usually organized in one equation \beqn \left(\begin{array}{c}
\boldsymbol{J}^{e}\\
\boldsymbol{J}^{h}
\end{array}\right)=\left(\begin{array}{cc}
\sigma & \alpha\\
T\alpha & \bar{\kappa}
\end{array}\right)\left(\begin{array}{c}
\boldsymbol{E}\\
-\nabla T
\end{array}\right).
\eeqn The electrical conductivity $\sigma$ and thermoelectric
transport coefficients $\alpha,\bar{\kappa}$ are matrices of
spatial coordinates. When $\boldsymbol{J}^{e}=0$, the thermal
conductivity
is given by $\kappa=\bar{\kappa}-T\alpha\sigma^{-1}\alpha$. 
The semiclassical equation of motion of partons in electric and
magnetic fields reads \beqn \dot{\boldsymbol{x}} &
\equiv\boldsymbol{v}_{n}(\boldsymbol{k})=\frac{1}{\hbar}\frac{\partial\epsilon_{n}(\boldsymbol{k})}{\partial\boldsymbol{k}}-\dot{\boldsymbol{k}}\times\boldsymbol{\Omega}_{n}(\boldsymbol{k}),\nonumber
\cr\cr \hbar\dot{\boldsymbol{k}} & =- e_\ast
\boldsymbol{E}(\boldsymbol{x}) - e_\ast
\dot{\boldsymbol{x}}\times\boldsymbol{B}(\boldsymbol{x}),
\label{eq: parton eom} \eeqn where $n$ is the band index, and
$\boldsymbol{\Omega}_{n}(\boldsymbol{k})$ is the Berry curvature
associated with each band.

We first evaluate the diagonal thermoelectric response by
neglecting the magnetic field $\boldsymbol{B}$ and the Berry
curvature $\boldsymbol{\Omega}_{n}$. With nonzero electric field
and temperature gradient, the solution of the Boltzmann equation
Eq.~\ref{eq: Boltzmann eq} for deconfined partons reads \beqn
\delta g=-\left(e_\ast
\boldsymbol{E}+\frac{\epsilon(\boldsymbol{k})-\mu}{T}\nabla
T\right)\cdot\boldsymbol{v}(\boldsymbol{k})\tau(\epsilon(\boldsymbol{k}))\left(-\frac{\partial
f}{\partial\epsilon}\right), \eeqn which leads to the diagonal
transport coefficients
\begin{flalign}
\sigma_{xx} & =e_{*}^{2}\mathsf{s}_{xx}(\epsilon_{F}^{*})\sim\frac{\mathsf{s}_{xx}(\epsilon_{F}^{*})}{N^{2}},\\
\alpha_{xx} & =-\frac{e_{*}}{T}\int d\epsilon\left(-\frac{\partial f}{\partial\epsilon}\right)(\epsilon-\mu)\mathsf{s}_{xx}(\epsilon)\sim\frac{T\mathsf{s}_{xx}^{\prime}(\epsilon_{F}^{*})}{N},\nonumber \\
\bar{\kappa}_{xx} & =\frac{1}{T}\int d\epsilon\left(-\frac{\partial f}{\partial\epsilon}\right)(\epsilon-\mu)^{2}\mathsf{s}_{xx}(\epsilon)\sim T\mathsf{s}_{xx}(\epsilon_{F}^{*}),\nonumber
\end{flalign}
where we have used $(-\partial
f/\partial\epsilon)\approx\delta(\epsilon-\epsilon_{F}^{*})$ and
$\mu\approx\epsilon_{F}^{*}$ at low temperature, and the function
$\mathsf{s}_{ij}(\epsilon)$ is defined as \beqn
\mathsf{s}_{ij}(\epsilon)=\tau(\epsilon)\int\frac{d^{2}\boldsymbol{k}}{(2\pi)^{2}}\delta(\epsilon-\epsilon(\boldsymbol{k}))v_{i}(\boldsymbol{k})v_{j}(\boldsymbol{k}).
\eeqn Assuming the band mass is isotropic and
$\boldsymbol{k}$-independent, one reproduces the Drude form for
partons $\mathsf{s}_{ij}=\delta_{ij}\tau n_{*}/m_{*}$. The
thermopower $Q$ (i.e., Seebeck coefficient) of the charge
fractionalized metal is given by: \beqn Q(T <
T_{1})=\frac{\alpha_{xx}}{\sigma_{xx}}=-N\frac{\pi^{2}}{3}\frac{k_{B}^{2}T}{e}\frac{\sigma'}{\sigma}.
\eeqn Note that the mean free path $l_{\textrm{m}}$ gets cancelled
in the ratio.


Experimentally, one clear signature for a charge fractionalized
metal is the strong violation of the Wiedemann-Franz law. The
Lorentz number acquires a large factor due to charge
fractionalization: \beqn L(T <
T_{1})=\frac{\kappa_{xx}}{T\sigma_{xx}}=N^{2}\frac{\pi^{2}}{3}\frac{k_{B}^{2}}{e^{2}}.
\label{eq:WFlaw} \eeqn This strong violation of the
Wiedemann-Franz law can be naively understood by the fact that,
though each fermionic parton carries a much smaller charge, it
still carries the same entropy as an electron.

When the temperature reaches $T_2$ and the partons are fully
confined, we expect these transport coefficients to decrease due
to confinement \beqn \frac{Q(T_{2})}{Q(T <
T_{1})}\sim\frac{1}{N},\qquad\frac{L(T_{2})}{L(T <
T_{1})}\sim\frac{1}{N^{2}}. \label{eq: ratios Seebeck and Lorentz}
\eeqn


For systems that break time-reversal symmetry such as a
ferromagnetic metal, the transport coefficients
$\sigma,\alpha,\kappa$ could have nonzero off-diagonal terms even
in the absence of $\boldsymbol{B}$. They receive intrinsic
contributions from the Berry curvature in the band structure.
Considering the nonzero Berry curvature
$\boldsymbol{\Omega}(\boldsymbol{k})$ in Eq.~\ref{eq: parton eom},
the parton wave packet acquires an anomalous velocity orthogonal
to $\boldsymbol{E}$, which leads to the anomalous Hall
conductivity \beqn
\sigma_{xy}(\epsilon)=\frac{e_{*}^{2}}{\hbar}\int\frac{d^{2}\boldsymbol{k}}{(2\pi)^{2}}\Theta(\epsilon-\epsilon(\boldsymbol{k}))\Omega^{z}(\boldsymbol{k}),
\eeqn where $\Theta(\epsilon)$ is the Heaviside step function.
Similar to their diagonal counterparts, the thermal Hall
conductivity $\kappa_{xy}$ is given by \beqn
\kappa_{xy}(\epsilon)=\frac{\pi^{2}}{3}\frac{k_{B}^{2}T}{e_{*}^{2}}\sigma_{xy}(\epsilon).
\eeqn At low temperature, the transverse Wiedemann-Franz law is
still strongly violated due to charge fractionalization.

\section{Fractionalized metal with $\SU(N)$ gauge field}

In this section we consider a more complex example of metal with
charge fractionalization. For simplicity we will consider spin
polarized electrons, hence the electron operator no longer carries
a spin index. The first step of our construction is a parton
construction: \beqn c_j \sim  \sum_{\{\alpha_i\}}
\epsilon_{\alpha_1, \alpha_2, \cdots \alpha_N} f_{j, \alpha_1}
f_{j, \alpha_2} \cdots f_{j, \alpha_N}. \label{eq: parton
construction} \eeqn This parton construction introduces a $\SU(N)$
gauge degree of freedom with an odd integer $N$. The nonabelian
gauge field was also first introduced for particle
physics~\cite{yangmills}, but later used broadly in the study of
spin liquids (see for example Ref.~\onlinecite{wenspinliquid2}),
and other strongly correlated electron
systems~\cite{chargeliquid,subirxumajorana}. The parton $f_\alpha$
with $\alpha = 1 \cdots N$ carries a fundamental representation of
the $\SU(N)$ gauge group, and also physical electric charge
$e_*=e/N$. The starting point of our analysis is the following
Lagrangian:
\begin{align}
&\mathcal{L}_{\rm UV}\left[\Psi^\dg, \Psi, a \right]=\mathcal{L}_{{\rm UV},\left[ \Psi, a \right]}+\mathcal{L}_{{\rm UV}, a}\nn\\
&\mathcal{L}_{{\rm UV},\left[ \Psi, a \right]}=\Psi^{\dg}\left(\ii \partial_t + g a_0^{I} t^I + e_* A_0+\mu  \right)\Psi \nn\\
&\qquad\qquad\quad-\frac{1}{2 m}\Psi^\dg\left(-\ii \partial_i + g a^I_i t^I+e_* A_i \right)^2\Psi\nn\\
&\mathcal{L}_{{\rm UV}, a} = -\frac{1}{4}F^I_{\mu \nu} (F^I)^{\mu
\nu} \label{eq:Luv}
\end{align}
where $\Psi=( f_1,f_2, ..., f_N )^t$, $a^I_\mu= (a^I_0, \
\vec{a}^I)$ is the $\SU(N)$ gauge field with $\mu=0,1,2$; $t^I$ is
the $\SU(N)$ Lie algebra in the fundamental representation, with
$I = 1, 2, ..., N^2-1$. $g$ is the strength of the gauge coupling,
the non-Abelian gauge field stress tensor is
$F^I_{\mu\nu}=\partial_\mu a_\nu^I - \partial_\nu a_\mu^I +
g\mathsf{\epsilon}^{IJK}a^J_\mu a^K_\nu$; $A_\mu = (A_0, \
\vec{A})$ is the background $\U(1)$ electromagnetic field.

Just like all systems that involve nonabelian gauge fields,
Eq.~\eqref{eq:Luv} needs gauge fixing. The systematic method of
gauge fixing is through the Faddeev-Popov
procedure~\cite{faddeev}, by introducing the ghost fields. Since
our system does not have Lorentz invariance to begin with, we will
consider the Coulomb gauge $\bm \nabla\cdot \vec{a}=0$. It was
shown in Ref.~\onlinecite{Zou2020} that the ghost fields are
decoupled from the system in the infrared limit. Further more, the
nondynamical component of the gauge field $a_0$ is suppressed by
the Thomas-Fermi screening of the Fermi surface, hence will be
dropped in the rest of the consideration.

Eq.~\ref{eq:Luv} with a finite Fermi surface is a highly
challenging theory to study. Starting with the Lagrangian
Eq.~\ref{eq:Luv}, one standard approximate treatment is to
construct the low energy effective theory assuming that the Fermi
energy is the largest energy scale in the problem, and $T \ll
E_F$. Below the cutoff $\Lambda$ that satisfies $T \ll \Lambda \ll
E_F$, the fermion operators can be expanded on two opposite
patches of the Fermi surface. A patch model can be constructed
following the logic of
Ref.~\onlinecite{polchinskinfl,nayaknfl1,nayaknfl2,nfl2,Mross2010,nfl4,nfl5,nflpairing1,Zou2020}.
The patch lagrangian $\mathcal{L}_{{\rm patch},\left[ \Psi, a
\right]}$ reads \beqn \mathcal{L}_{{\rm patch},\left[ \Psi, a
\right]}&=&\Psi^\dg \left[\ii \eta \partial_t -(l v_F
k_x+\frac{k_y^2}{2m})\right]\Psi \cr\cr &-& l g v_F \Psi^\dg a_x^I
t^I \Psi\nn ; \cr\cr \mathcal{L}_{{\rm patch},a} &=&
\frac{1}{2}q_y^2 ( a_x^I)^2. \label{eq:Lpatch} \eeqn $x$ and $y$
are the local coordinates orthogonal and transverse to the patch
Fermi surface of interest, $l = \pm 1$ labels the antipodal
patches that can be connected by the transverse gauge
fluctuations. The form of the patch Lagrangian implies the
following scaling of space-time coordinates under coarse graining
\beq \omega'=b^{z_f} \omega, \ k'_x=b k_x, \ k'_y=b^{1/2} k_y \eeq
with $z_f=1$, and $b>1$. Due to the different scaling dimensions
of the $x$ and $y$ coordinate and the Coulomb gauge constraint, we
find $\Delta_{a_y} = \Delta_{a_x} + 1/2$, where
$\Delta_{\mathcal{O}}$ is the scaling dimension of a field or
coupling $\mathcal{O}$, e.g.\ $a'_y = b^{\Delta_{a_y}} a_y$, $a'_x
= b^{\Delta_{a_x}} a_x$. Due to the highly anisotropic scaling of
space-time, the form of the Lagrangian of the patch theory
Eq.~\ref{eq:Lpatch} is very different from a standard Lorentz
invariant theory.

Unlike the $\U(1)$ gauge theory, a non-Abelian gauge field has
self-interactions. It can be argued within the framework of the
patch theory~\cite{Zou2020} that the self-interaction between
gauge bosons is irrelevant in the infrared, hence we can use
Eq.~\ref{eq:Lpatch} as the starting point of RG analysis. Note
that the irrelevance of gauge field self-interactions is due to
the highly anisotropic scaling of local coordinates $x$ and $y$ in
Eq.~\ref{eq:Lpatch}. To get a controlled interacting RG fixed
point, we need one more step of transformation of
Eq.~\ref{eq:Lpatch}: we consider a small $\epsilon$ expansion by
replacing $q_y^2$ with $q_y^{1+\epsilon}$, as was first introduced
by Ref.~\onlinecite{nayaknfl1,nayaknfl2}. At the leading order of
$\epsilon$, only the 1-loop diagrams contribute, which leads to a
weakly interacting RG fixed
point~\cite{nayaknfl1,nayaknfl2,Mross2010}. The self-energy
correction to the parton propagator $\mathrm{G}=-\ii \langle T_t
\Psi \bar{\Psi}\rangle$ obtained by integrating out modes from
$q_y=\Lambda$ to $q_y=\Lambda/b^{1/2}$ reads
\begin{align}
\delta \Sigma=-\ii \omega \frac{g^2 v_F}{4\pi^2 }\mathbb{c}_2
\boldsymbol{1} \ln b \label{wf}
\end{align}
where $\mathbb{c}_2 \bs{1}=\sum_I t^I t^I =\frac{N^2-1}{2N}
\bs{1}$, with $\mathbb{c}_2$ the quadratic Casimir operator for
the fundamental representation of $\SU(N)$; and $\boldsymbol{1}$
is the identity matrix in the color space. The vertex correction
vanishes at the leading order $\epsilon$-expansion, as was argued
in Ref.~\onlinecite{Mross2010}. Eventually the one-loop
corrections lead to a new fixed point $g_*^2= 2\pi^2 \epsilon /
(\mathbb{c}_2 v_F)$. The existence of this fixed point is
physically due to the screening of the gauge coupling from matter
fields with finite density of states.

Physical properties at this new fixed point can be
self-consistently solved. To be general, we consider the gauge
field kinetic energy as $k_y^2\rightarrow |k_y|^{1+\epsilon}$,
while $\epsilon$ is not necessarily small for the self-consistent
calculation. Assuming that the parton self-energy does not depend
on the momentum, which can be checked posteriori, the
self-consistent equation reads
\beqn && \Sigma_{\alpha \alpha'}(\ii \omega,\boldsymbol{k}) =
\sigma_f (\ii \omega,\boldsymbol{k}) \delta_{\alpha \alpha'};
\cr\cr && \sigma_f (\ii \omega,\boldsymbol{k}) = (-) g^2 v_F^2
\mathbb{c}_2 \cr\cr &\times& \int \frac{\dd \nu \dd q_x \dd
q_y}{(2\pi)^3} \frac{1}{\ii (\omega+\nu)
+\sigma_f(\omega+\nu)-\xi_{\bs{k}+\bs{q}}} \cr\cr &\times&
\frac{1}{\ii \nu +\pi_a(\ii \nu ,\bs{q})+q_y^{1+\epsilon}} ;
\cr\cr\cr && \Pi_{I J}(\ii \nu ,\bs{q}) = \pi(\ii \nu ,\bs{q})
\delta_{IJ}; \cr\cr && \pi(\ii \nu ,\bs{q}) = g^2 v_F^2 \mathbb{c}
\int \frac{\dd \bs{k} \dd \omega}{(2\pi)^3} \frac{1}{\ii
\omega+\sigma_f(\omega)-\xi_{\bs{k}}} \cr\cr &\times& \frac{1}{\ii
(\omega+\nu) +\sigma_f(\omega+\nu)-\xi_{\bs{k}+\bs{q}}}. \eeqn
The solution for the fermion and gauge boson self-energy is \beqn
\mathrm{\Sigma}_{\alpha \alpha'}(\omega) &=& -\ii
\frac{\mathbb{c}_2 \mathbb{c}^{-\epsilon/(2+\epsilon)}}{2
(8\pi)^{\epsilon/(2+\epsilon)}} \csc (\frac{2\pi}{2+\epsilon})\,
\bar{g}^{\frac{2}{2+\epsilon}} E_f^{\frac{\epsilon}{2+\epsilon}}
\cr\cr &\times& |\omega|^{\frac{2}{2+\epsilon}} \sgn(\omega)
\delta_{\alpha \alpha'}\nn \cr\cr \pi(\ii \nu,\boldsymbol{q}) &=&
\frac{1}{v_F^{2+\epsilon}} \frac{\mathbb{c}}{8\pi} \bar{g}
E_f^{1+\epsilon} \left|\frac{\nu}{\boldsymbol{q}}\right|. \eeqn
Here we have defined a dimensionless coupling constant
$\bar{g}=\frac{g^2 v_F^{1-\epsilon}}{(2m)^{\epsilon}}$. One can
see that the standard Landau damping term emerges in the
self-consistent solution of the gauge boson self-energy. And the
fermion self-energy takes the form of a non-Fermi liquid.

$ $

{\it --- Confinement and Crossover at finite temperature}

To evaluate transport properties at different temperature scales,
like the $Z_N$ gauge theory discussed earlier, we need to
determine the two temperature scales $T_1$ and $T_2$ at which the
confinement length satisfies $\xi_c(T_1)\sim l_\mr$ and
$\xi_c(T_2)\sim a$. When $\xi_c > l_\mr$ the transport is governed
by fractionalized charges. Like the case with the $Z_N$ gauge
field, here we need to evaluate the scaling of $\xi_c$ with
temperature at the fixed point discussed above, and in this
section we are going to take $\epsilon = 1$. First of all, the
gauge fields would become classical when $g_\ast^2 \frac{|\nu_{n=
1}|}{q}> q^2$, where $\nu_n$ is the $n$-th Matsubara frequency.
This gives a quantum-classical crossover length $\xi_{\rm cl}\sim
q_{\rm cl}^{-1} \sim (T g_\ast^2)^{-\frac{1}{3}}$ above which the
gauge field dynamics is classical. A classical gauge theory in
$2d$ is described by the action
\begin{align}
\mathcal{S}_{\rm classical}=\int \dd \bs{x} \ \sum_I \frac{1}{T
g^2 } (F^I_{\mu\nu})^2
\end{align}
The scaling dimension of $T g^2$ now becomes $\Delta_{T g^2}= 2$.
At the confinement length $\xi_c$, $T g^2$ renormalizes to $T g^2
\sim 1$. We then find \beqn && \frac{T g(\xi_c)^2}{T g_\ast^2}
\sim \frac{1}{T g_\ast^2} \sim \left(\frac{\xi_c}{\xi_{\rm
cl}}\right)^{2} \cr\cr &\rightarrow& \xi_c(T)\sim  (T
g_\ast^2)^{-5/6}\sim T^{-5/6}. \eeqn Hence at low temperature $T$,
due to the Landau damping physics arising from the Fermi surface,
when we observe the system with increasing length scale, physics
of the gauge field will first crossover to classical at $\xi_{\rm
cl} \sim (T g_\ast^2)^{-1/3}$, then crossover to confinement at an
even longer scale $\xi_c(T)\sim (T g_\ast^2)^{-5/6}$. This
analysis implies that the crossover temperature $T_1$ scales with
the mean free path $T_1 \sim l_\mr^{-6/5}$.

$ $

{\it --- Transport Properties}

At low temperature, we assume that the impurity still dominates
the momentum relaxation. This assumption is valid at strictly zero
temperature, and also valid at finite temperature with the
artificial limit of small $\epsilon$, since the fixed point gauge
coupling $g_\ast^2 \sim \epsilon$, scattering with the gauge field
is weak in this limit. The resistivity caused by gauge boson
scattering can be computed following the procedure in
Ref.~\onlinecite{leenagaosa}. One key difference from the $Z_N$
example we discussed before is that, there are $N$ species of the
fermionic partons now, each with the same density as the electron,
and hence the same size of Fermi sea as the electron, $i.e.$
$n_\ast = n_e$, and $k_{F}^\ast=k_F$. In this case, the
conductivity of the fractionalized metal at zero temperature reads
\beqn \sigma(T=0)= N \left( \frac{n_\ast e_\ast^2 l_\mr}{m_\ast
v^\ast_F} \right) \sim \frac{1}{N} \frac{e^2}{h} (l_\mr
n_e^{1/2}), \eeqn which can still be a bad metal. Notice that in
other parton constructions for example in
Ref.~\onlinecite{senthilmit1}, the total electrical conductivity
is governed by the Ioeffe-Larkin rule~\cite{larkin}; while in our
case the conductivity should be a direct sum of conductivity of
each parton. Once again, when the confinement length $\xi_c$
becomes the order of lattice constant $a$ (which occurs at
temperature $T_2$ with $\xi_c(T_2) \sim a$ ), the partons are
fully confined to electron, and the conductivity is given by the
standard form $\sigma(T_2) = \frac{e^2}{h} (l_\mr n_e^{1/2})$.

At low temperature, both the partons and the gauge bosons will
contribute to the thermal transport. But it was shown that the
gauge boson contribution is subdominant~\cite{navelee} compared
with the fermionic partons,
hence it will be ignored in the following discussion. At low
temperature the thermal transport of the fermionic partons will
also be mostly determined by their scattering with impurities:
\beqn \frac{\kappa}{T} = N \left( \frac{\pi^2}{3} \frac{k_B^2
n_\ast l_\mr}{m_\ast v_F^\ast} \right) \sim N \frac{\pi^2}{3}
\frac{k_B^2}{h} (l_\mr n_e^{1/2}), \eeqn again we have taken into
account of the fact that, there are $N$ color species of the
partons, and for each species $n_\ast = n_e$. There is still a
strong violation of the Wiedmann Franz law same as the $Z_N$ gauge
field case Eq.~\eqref{eq:WFlaw} at zero temperature:
\begin{equation}
L(T=0) = \frac{\kappa_{xx}}{T\sigma_{xx}} =
N^{2}\frac{\pi^{2}}{3}\frac{k_{B}^{2}}{e^{2}}.
\end{equation}

\section{Summary and Discussion}

We proposed two constructions of exotic metallic phases based on
the idea of charge fractionalization. It was proposed before that
charge fractionalization may be playing an important
role~\cite{xumit} in the metal-insulator transition (MIT) observed
in transition metal dichalcogenide (TMD) moir\'{e}
heterostructures~\cite{tmdmit1}, where an anomalously large
resistivity was observed at low temperature near and at the MIT,
followed by a rapid drop of resistivity at slightly higher
temperature, analogous to the physics discussed between $T_1$ and
$T_2$ in Fig.~\ref{resistivity}. Similar physics has also been
observed in another TMD moir\'{e} sample~\cite{tmdmitwu}.

The two constructions discussed in this work are actually related
to each other. The $\SU(N)$ gauge group always has a $Z_N$ center,
hence the $\SU(N)$ gauge field can be broken down to a $Z_N$ gauge
field by condensing Higgs fields~\cite{higgs1,higgs2,higgs3} with
the right representation. The condensed Higgs field is also
expected to mix the different color species and lift the
degeneracy of the fermionic parton Fermi surface. In fact, a spin
liquid usually has a $\U(1)$ or even $\SU(2)$ gauge degrees of
freedom in the UV, but the gauge group can be broken down to $Z_2$
through the Higgs mechanism, hence in the infrared the system
becomes a $Z_2$ spin
liquid~\cite{wenspinliquid1,subirspinliquid,wenspinliquid2}.

Xu's group is supported by NSF Grant No. DMR-1920434, and the
Simons Investigator program; Z.L.is supported by the Simons
Collaborations on Ultra-Quantum Matter, grant 651440 (LB); M.Y.
was supported in part by the Gordon and Betty Moore Foundation
through Grant GBMF8690 to UCSB, and by the NSF Grant No.
PHY-1748958. The authors thank Chao-Ming Jian for very helpful
discussions.

\bibliography{metal}

\end{document}